\documentclass[
12pt
  ]
 {article}

\usepackage{a4wide}
\usepackage{bm}

\def\e{\varepsilon}

\def\ra{\rightarrow}

\begin{document}

\title{Turbulence models of gravitational clustering}



\author{Jos\'e Gaite\\
{\small IDR, ETSI Aeron\'auticos, Universidad Polit\'ecnica de Madrid,}\\ 
{\small Pza.\ Cardenal Cisneros 3, E-28040 Madrid, Spain}
}

\date{February 15, 2012}

\maketitle

\begin{abstract}
Large-scale structure formation can be modeled as a nonlinear process that
transfers energy from the largest scales to successively smaller scales until
it is dissipated, in analogy with Kolmogorov's cascade model of incompressible
turbulence. However, cosmic turbulence is very compressible, and vorticity
plays a secondary role in it.  The simplest model of cosmic turbulence is the
adhesion model, which can be studied perturbatively or adapting to it
Kolmogorov's non-perturbative approach to incompressible turbulence.  This
approach leads to observationally testable predictions, e.g., to the power-law
exponent of the matter density two-point correlation function.
\end{abstract}

\section{Introduction}

The evolution of the Universe is ruled by gravity.  Although the Einstein
equations of General Relativity (GR) are strongly nonlinear, the early
evolution of the Universe can be described by a FLRW solution plus {\em
  linear} perturbations.  Since the FLRW solutions are unstable, perturbations
grow, and they grow faster on smaller scales, becoming nonlinear and
decoupling from the global expansion.  As instability is an inherent property
of self-gravitating systems, they are bound to cluster and collapse as long as
they can dissipate their kinetic energy.  This dissipation is essentially a
transfer of kinetic energy from one scale to another smaller scale, due to
nonlinear mode coupling.  The kinetic energy eventually reaches the smallest
scales and disappears (as discussed in page \pageref{dissip}).

The nonlinear transfer of kinetic energy from larger to smaller scales is the
hallmark of fluid turbulence and is the basis of Kolmogorov's cascade model of
incompressible turbulence, namely, of turbulent solutions of the Navier-Stokes
equations with large Reynolds number but small Mach number \cite{Frisch}.  In
cosmology and, in particular, in cold-dark-matter (CDM) models, the Mach
number of the matter fluid has to be large, so the turbulence is actually very
compressible. In contrast, the fluid velocities are non-relativistic (except
on very small scales). Therefore, a Newtonian treatment is generally
appropriate. In fact, the full Newtonian dynamics is still too complex for an
analytical approach, so it is further simplified as follows.

\section{Zeldovich approximation and adhesion model}

The Newtonian equation of motion of a test particle in an expanding background
is best expressed in terms of the comoving coordinate ${\bm x} = {\bm r}/a(t)$
and the {\em peculiar} velocity ${\bm u} = \dot{\bm x} = {\bm v} - H {\bm r}$.
So $d {\bm v}/{d t} = {\bm g}_\mathrm{T}$ is written as $d {\bm u}/{d t} + H
{\bm u} = {\bm g}$, in terms of the peculiar gravitational field, ${\bm g} =
{\bm g}_\mathrm{T} - {\bm g}_\mathrm{b}$, with background field ${\bm
  g}_\mathrm{b}=\dot{H} {\bm r} + H^2{\bm r}$, such that $3(\dot{H} + H^2) =
\nabla\cdot{\bm g}_\mathrm{b} = -4\pi G\rho_\mathrm{b}\,,$ which is just the
dynamical FLRW equation for $P=0$ (the dust model). Of course, ${\bm g}$
depends on the motion of the remaining particles, and the problem is
nonlinear, like in GR. It can be linearized when the peculiar variables are
small, so the motion is, approximately, ${\bm x} = {\bm x}_0 + b(t)\,{\bm
  g}({\bm x}_0)$, where $b(t)$ is the growth rate of linear density
fluctuations. Redefining time as $\tau=b(t)$, the motion is simply linear
motion, with a constant velocity given by the initial peculiar gravitational
field. Naturally, nearby particles have different velocities, and, as the
linear solution is prolonged into the nonlinear regime, trajectories cross at
{\em caustic} surfaces \cite{Shan-Zel}, \label{caustic} called ``Zeldovich
pancakes'' in this context. These are supposed to be the first cosmological
structures.

Actually, the simplest caustic arises from inward spherical motion (spherical
collapse), which gives rise to a point singularity, that is, a
zero-dimensional (degenerate) caustic.  After the multiple collision of
particles at one point, their evolution in Newtonian gravity is undefined. If
no kinetic energy is dissipated ({\em adiabatic} collapse), the particles
cross (or rebound), returning to their initial positions. This unrealistic
evolution takes place at any caustic, so there is no real structure formation,
unless a dissipation mechanism is introduced. Hence, the linear motion in the
Zeldovich approximation is supplemented with a {\em viscosity} term, resulting
in the equation
\begin{equation}
\frac{d \widetilde{\bm u}}{d \tau} \equiv
\frac{\partial \widetilde{\bm u}}{\partial \tau} + 
\widetilde{\bm u}\cdot \nabla\widetilde{\bm u} =
\nu \nabla^2\widetilde{\bm u}, 
\label{Burg}
\end{equation}
where $\widetilde{\bm u}$ is the peculiar velocity in $\tau$-time.  To this
equation, it must be added the no-vorticity (potential flow) condition,
$\nabla \times \widetilde{\bm u} = 0$, implied by $\nabla \times {\bm g}({\bm
  x}_0) = 0$.  Eq.~(\ref{Burg}) is the Burgers equation for very compressible
(pressureless) fluids. The limit ${\nu \ra 0}$ might seem to recover the
caustic-crossing solutions but actually is the high Reynolds-number limit and
gives rise to Burgers turbulence. Whereas incompressible turbulence is
associated to the development of vorticity, Burgers turbulence is associated
to the development of {\em shock fronts}, namely, discontinuities of the
velocity.  These discontinuities arise at caustics and give rise to matter
accumulation by inelastic collision of particles.  The viscosity $\nu$
measures the thickness of shock fronts, which become true singularities in the
limit ${\nu \ra 0}$.  This is the {\em adhesion model}, which produces a
characteristic network of sheets, filaments and nodes, called ``the cosmic
web.''

\section{Approaches to cosmic Burgers turbulence}

Unlike the Navier-Stokes equation of incompressible fluids, the Burgers
equation is {\em integrable} and, therefore, keeps memory of initial
conditions. On the other hand, in both incompressible and Burgers turbulence,
kinetic energy cascades down to smaller scales until it is dissipated.  This
dissipation is of thermal nature in normal fluids; \label{dissip} but, in
cosmology, while baryons experience thermal dissipation, the ``dissipation''
in CDM has mainly gravitational origin \cite{Gur-Zyb}.  Therefore, it is not
sensible to just neglect the dissipated energy in Eq.~(\ref{Burg}).  Moreover,
the gravitational equations of motion are {\em chaotic}, like the
Navier-Stokes equation, and tend to lose memory of initial conditions.  In
Burgers turbulence, we can recover the dissipated energy {\em and} have a
stationary state with no memory of initial conditions (a fractal attractor,
actually) by adding a ``noise'' to Eq.~(\ref{Burg}), giving rise to the {\em
  stochastic adhesion model}.  A white noise is appropriate for thermal
fluctuations, but we must allow for non-thermal fluctuations and also take
into account the large-scale energy pumping. Both can be modeled together as
some ``colored'' stochastic forcing.  Therefore, the total random force, which
must derive from a potential, ${\bm f}= \nabla \eta$, is just assumed to be
Gaussian, with zero mean and white-in-time covariance:
\begin{equation}
\langle \eta({\bm x},t)\,\eta({\bm x}',t')\rangle =  D({\bm x}-{\bm
  x}')\,\delta(t - t'). 
\label{noise}
\end{equation}
This stochastic adhesion model correctly describes the cosmic web structure
as a ``quasi-Voronoi'' tessellation of shock fronts \cite{Molchan}.

Let us see, first, a perturbative approach to this model and, second, a
non-perturbative one, based on the Kolmogorov approach to incompressible
turbulence \cite{TMCS}.

\subsection{Perturbative approach}

The stochastic Burgers equation has been studied in a very different context,
namely, the statistical description of surface growth: when surface's height
is identified with velocity potential, the KPZ equation for surface growth is
equivalent to the stochastic Burgers equation \cite[p.\ 61]{Bar-Stan}.  It can
be studied with perturbation theory, which shows that a dynamical scaling
appears at the fixed points of the dynamical renormalization group
\cite{Bar-Stan}.  Unfortunately, the nontrivial fixed point is repulsive (in
three spatial dimensions), so the nonlinear term of the Burgers equation is
{\em irrelevant} (in the renormalization-group sense) and the viscous term
dominates in the perturbative stationary state.  Therefore, turbulence can
only occur in the strong-coupling, non-perturbative regime.

In fact, the effective coupling constant in the renormalization group
equations has the generic expression $\lambda^2D/\nu^3$ \cite{Bar-Stan}, where
$\lambda$ is a coupling constant for the nonlinear term in Eq.~(\ref{Burg}),
$D$ is the noise strength, as in Eq.~(\ref{noise}), and $\nu$ is the
viscosity. As turbulence occurs when $\nu\ra 0$, the coupling must be
strong. More precisely, the effective coupling constant is actually
proportional to the cube of the Reynolds number, which has to be very large,
making perturbation theory unreliable.

\subsection{Kolmogorov like approach}

A non-perturbative approach can be based on Kolmogorov's universality
assumptions, namely, statistical homogeneity, isotropy, and velocity scaling
laws \cite{Frisch}.  These laws stem from the principle that the only
parameter relevant for the turbulent cascade is the energy flux across the
scales, per unit time and per unit mass, $\e$. Therefore, velocity correlation
functions, for example, must be {\em power-laws}, with exponents determined by
dimensional analysis; in particular,
$$
\langle \widetilde{\bm u}({\bm x}) \,\widetilde{\bm u}({\bm x}')   \rangle \propto
(\e |{\bm x} - {\bm x}'|)^{2/3} .
$$
From this expression and the relation $\nabla\cdot\widetilde{\bm u} \propto
\nabla\cdot{\bm g} \propto \delta\rho = \rho-\rho_\mathrm{b},$ we can deduce the
matter-density reduced two-point correlation function:
\begin{equation}
\langle \delta\rho(\bm{x}) \, \delta\rho(\bm{0}) \rangle/\rho_\mathrm{b}^2 \propto
\langle \nabla\cdot\widetilde{\bm u}({\bm x})  \,\nabla\cdot\widetilde{\bm
  u}({\bm 0})   \rangle \propto 
|{\bm x}|^{-4/3}.
\label{2p-correl}
\end{equation}
In fact, the preceding relation between $\widetilde{\bm u}$ and $\rho$ is only
approximate and, furthermore, one should take account of {\em intermittency},
which causes deviations from Kolmogorov's scaling laws \cite{Frisch}.
Intermittency is due to spatial variations of $\e$, which are especially
strong in Burgers turbulence, because dissipation takes place in caustics.  A
detailed analysis \cite{TMCS} shows that Eq.~(\ref{2p-correl}) holds, but the
power-law exponent absolute value can be in the range $(1,4/3)$, depending on
the noise strength.  These values are smaller than the observational value,
about 1.7 \cite{Jones-RMP}.

\section{Discussion}

The derivation of the adhesion model involves drastic simplifications, and one
could object to Eq.~(\ref{Burg}) that it has almost lost track of gravity,
which only survives in the definition of $\tau$.  However, the essential feature
of the adhesion model is that it gives rise to caustics, and this is a general
feature of the irrotational motion of pressureless matter in Newtonian
dynamics or in GR (the significance and genericity of caustic singularities in
GR are discussed by, e.g., Landau and Lifshitz \cite{LL}).  Since GR is
nonlinear, just the definition of sheet, filament, and node singularities
demands a special study \cite{Ge-Trasch}.  Nodes, as point singularities, are
the strongest singularities and must give rise to black holes.  A black hole
has (quantum) internal configurations and {\em entropy}, which reflect the
energy dissipated in its formation. Moreover, black holes have {\em negative}
specific heat (like Newtonian self-gravitating systems) and cannot reach
thermal equilibrium in an unbounded environment \cite{Hawk}, so they have to
keep growing.  Black-hole growth involves a ``mixmaster'' dynamics that
exemplifies the ``gravitational turbulence'' of general solutions of the
Einstein equations \cite{Barrow}.  The formation and growth of super-massive
black holes are probably the most dissipative processes in astrophysics and
surely have contributed greatly to shape our present universe.

In any case, the adhesion model, in which all these small-scale processes are
lumped into an effective viscosity, seems to be adequate to describe the
formation of large-scale structure, at least, in an early stage and
semiquantitatively. The construction of more realistic models of cosmic
turbulence, whether Newtonian or relativistic, demands further theoretical
developments.

\end{document}